# Investigations into the complete spreading dynamics of a viscoelastic drop on a spherical substrate

*Sudip Shyam[1,$], Harshad Sanjay Gaikwad[1,$], Syed Abu Ghalib Ahmed[2], Bibek Chakraborty[2], Pranab Kumar Mondal\*[1].*

[1]*Microfluidcs and Microscale Transport Processes Laboratory,*
*Department of Mechanical Engineering, Indian Institute of Technology Guwahati, Assam,*
*India – 781039*

[2]*Department of Mechanical Engineering, Tezpur University, Napaam, Assam,*
*India – 781048*

[$]Authors have equal contribution

---

[\*]Corresponding author email address: mail2pranab@gmail.com, pranabm@iitg.ernet.in

**Abstract**

We study the spreading dynamics of a sphere-shaped elastic non-Newtonian liquid drop on a spherical substrate in the capillary driven regime. We use the simplified Phan–Thien–Tanner model to represent the rheology of the elastic non-Newtonian drop. We consider the drop to be a crater on a flat substrate to calculate the viscous dissipation near the contact line. Following the approach compatible with the capillary-viscous force balance, we establish the evolution equation for describing the temporal evolution of the contact line during spreading. We show that the contact line velocity obtained from the theoretical calculation matches well with our experimental observations. Also, as confirmed by the present experimental observations, our analysis deems efficient to capture the phenomenon during the late-stage of spreading for which the effect of line tension becomes dominant. An increment in the viscoelastic parameter of the fluid increases the viscous dissipation effect at the contact line. It is seen that the higher dissipation effect leads to an enhancement in the wetting time of the drop on the spherical substrate. Also, we have shown that the elastic nature of fluid leads to an increment in the dynamic contact angle at any temporal instant as compared to its Newtonian counterpart. Finally, we unveil that the phenomenon of increasing contact angle results in the time required for the complete wetting of drop becomes higher with increasing viscoelasticity of the fluid. This article will fill a gap still affecting the existing literature due to the *unavailability of experimental investigations of the spreading of the elastic non-Newtonian drop on a spherical substrate*.

## I. Introduction

Spreading of liquid drops on solid surfaces is encountered frequently in several natural phenomena such as falling of raindrops in the glass window, movement of drop on lotus leaves to different industrial applications. Among the second, a window of practical applications falls in this paradigm are the coating of a solid object by liquid layers, printing technology, micromixing, painting, and many more.[1–3] A wide gamut of practical applications alongside a rich physics involved with the underlying spreading phenomenon has motivated researchers to intensely concentrate as well as systematically interrogate several aspects of the spreading dynamics of liquid drop over solid substrates in recent years.[1,2,4–7] The dynamical behaviour of the drop spreading on a solid substrate is indeed fascinating in the capillary driven regime, attributed mainly to the complex interplay between surface tension modulated alteration in contact line velocity and the dissipative energy at the three-phase contact line. This phenomenon becomes further involved under the modulation of surface energy as controlled by the substrate wetting characteristics.[1,2,8–10]

The underlying physical issues concerned with the spreading of non-Newtonian drops on a solid substrate are more complex. The primary attributable factor behind this facet is the rheology driven modifications of the flow dynamics and its subsequent effect on the contact line motion.[11–13] Taking a note on this aspect, efforts have been directed at the exploration of complicated phenomenon delimited with the spreading of non-Newtonian drops on substrates having different geometrical topographies in recent times.[1,5,14–17] It is important to mention here that the underlying phenomenon of drop spreading on the spherical substrates is markedly different from that on the flat substrates, attributable to the intensified effect of *line tension* on the spreading dynamics.[6,18–22] At low contact angle, the effect of line tension, however small it is, becomes vital on the spreading dynamics of drops on a spherical substrate. As shown by the researchers,[18] a small but non-zero finite magnitude of line tension is an indicative measure for the complete wetting state of drop on a spherical substrate.[18]

Majority of the literature referenced above have dealt with the drop spreading dynamics considering either Newtonian fluid or non-Newtonian fluid with inelasticity on surfaces having different morphologies.[1,5,14–17] Important to mention, the spreading phenomenon of the elastic non-Newtonian drop, precisely viscoelastic drop, on a spherical substrate is expected to encounter complicated dynamics on account of a number of factors. This complicated dynamical behaviour is primarily triggered by the correlative-cooperative effect of fluid viscoelasticity on the underlying flow dynamics. It is worth to add here that the spreading of viscoelastic drop on a spherical substrate in the presence of line tension effect, which is more useful for ensuring the complete wetting state of the drop (late-stage), becomes even more convoluted as the contact-line radius tends to vanish (onset point of singularity). Considering above all aspects, the spreading of elastic non-Newtonian drop on spherical substrate seems to be of vital importance from two different perspectives: first, because of its direct consequence to practical relevance including printing, coating, etc. Second, for the understanding of rich physics involved with the underlying phenomenon. Important to mention, this aspect has not been explored to date as apparent from a comprehensive review of the reported literature in this paradigm.

Here, we investigate the spreading dynamics of a viscoelastic drop on a spherical substrate following a complex theoretical analysis. Also, to substantiate the efficacy of our theoretical model, we perform experiments for the spreading of both viscoelastic and Newtonian drops in the regime compatible with the capillary-viscous force balance. Our theoretical results of the spreading behaviour of the viscoelastic drop, and also for the Newtonian drop in a limiting case, match well with the experimental observations. Most importantly, our theoretical model developed in predicting the spreading behaviour of the elastic non-Newtonian drop, which encompasses the class of shear-thinning fluid along with a particular case of the Newtonian drop, benchmarked with the experimental data will provide a basis for further investigations towards capturing intricate details of the drop spreading dynamics. *Quite non-intuitively, the theoretical framework developed in this study can accurately predict the behaviour during the late-stage of the spreading, as confirmed by the experimental evidence under identical conditions.*

**II. Spreading of viscoelastic drop on a spherical substrate**
**A. Mathematical modelling**

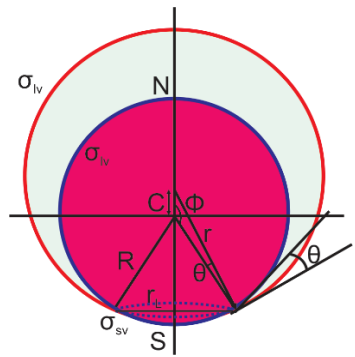

**Figure 1** (**colour online**): The plots depict the schematic of a spherical liquid drop spreading from the north pole 'N' of a spherical substrate towards its south pole 'S'. The distance between the center of the drop and the spherical substrate is given by $C$. Note that $r_L$ denotes the radius of the contact line, while θ is the dynamic contact angle, which is related to the half of the central angle φ.

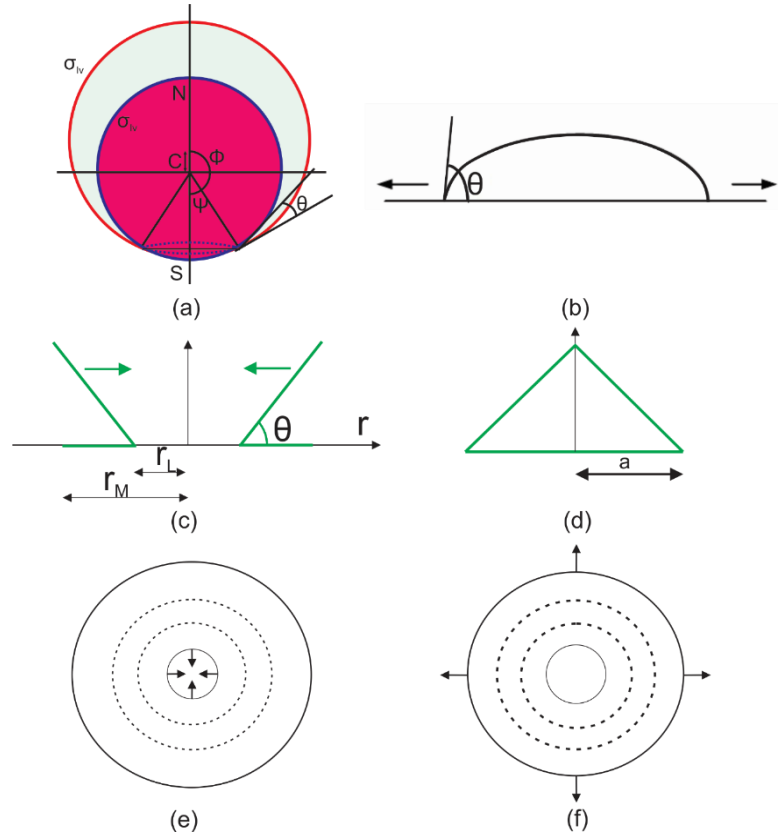

**Figure** 2 (**colour online**): (a) Schematic represents the spreading of a drop on a spherical substrate toward the complete-wetting state. At this stage, the three-phase contact line shrinks towards the south pole of the spherical substrate. (b) Plot depicts the drop spreading on a flat substrate. For this particular case, the three-phase contact line expands towards infinity. Thus, the line tension on a spherical substrate plays an opposite role to that on a flat substrate. (c) Schematic represents the crater-shaped model of spreading drop on a spherical substrate, while (d) shows the cone-shaped model of the drop on a flat substrate. (e) Plot depicts the bottom view of the spreading drop, whose contact line shrinks towards south pole $S$ of the spherical substrate, and (f) shows the top view of the spreading drop whose contact line expands towards infinity on a flat substrate.

We here consider the spreading of an elastic non-Newtonian drop on a spherical substrate as schematically shown in Fig. 1. For an accurate description of the rheology of the viscoelastic fluid, we use the simplified Phan-Thien-Tanner model. The simplified Phan-Thein-Tanner or sPTT model by its common name is derived on the basis of molecular theory, particularly the network theory, made of the polymeric strands.[23] Moreover, as can reviewed from the literature, this model appears to be the most commonly used one for describing the rheology of the viscoelastic fluids.[24–26] It may be mentioned here that this model can depict both the shear-thinning nature as well as the elastic nature of the viscoelastic fluids. Notably, in several experimental and computational studies, it has been shown that the sPTT model precisely mimics the flow behaviour of the shear-thinning viscoelastic fluids such as a solution of Polyisobutylene in Tetradecane (PIB/C14), Polyacrylamide (PAA), Polyethylene Oxide (PEO), and Low-Density Polyethylene (LDPE).[27–33] Therefore, it is easily perceptible that the sPTT model is deemed an effective model for analysing the flow behaviour of the shear-thinning viscoelastic fluid. On account of this significance, we use in the present study the sPTT model to depict the spreading phenomenon of the viscoelastic fluid on a spherical substrate. Following this model, we can write the constitutive equation for the viscoelastic fluid as [34,35]

$$f(\text{tr}(\boldsymbol{\tau}))\boldsymbol{\tau} + \lambda\left[\frac{\partial \boldsymbol{\tau}}{\partial t} + u.\nabla\boldsymbol{\tau} - \left((\nabla u)^T.\boldsymbol{\tau} + \boldsymbol{\tau}.\nabla u\right)\right] = 2\eta \mathbf{D} \qquad (1)$$

In Eq. (1), $u$ is the velocity field, $\boldsymbol{\tau}$ is the stress tensor, $\lambda$ is the relaxation time, $\mathbf{D}$ is the deformation rate tensor, $f(\text{tr}(\boldsymbol{\tau}))$ is a function of the trace of the stress tensor, and $\eta$ is the viscosity coefficient or zero-shear viscosity. Before proceeding to outline the detailed analysis of the problem considered in this study, we would like to mention here that the present analysis focuses on the late-stage spreading of the drop. The spreading of the drop on any surface can be modelled using the combination of the molecular kinetic theory (MKT) and the hydrodynamic approach.[5] Theoretical description of underlying events associated with the droplet spreading during early-stage needs employment of MKT (molecular kinetic theory) model in the underlying analysis.[36,37] On the other hand, the hydrodynamic approach can give accurate results for the late-stage spreading where both the line tension effect and the frictional effect due to viscous force are taken into account.[5] In the underlying analysis of the late-stage spreading, the effect of line tension develops when the contact angle tends to zero, thus signifying the complete wetting. In the present study, we have used the hydrodynamic approach consistent with the crater-shaped model to study the late-stage spreading of the viscoelastic fluid on the spherical substrate. Note that the crater-shaped model approximates the complex geometry of the drop spreading phenomenon on the spherical substrate. The late-stage spreading phenomenon of the viscoelastic drop on a spherical substrate by considering a shrinking crater on a flat substrate is shown through a series of schematic depictions in Fig. 2.[4,5] We can see from Fig. 1 that, as the drop is advancing, the three-phase contact line is shrinking towards the south pole 'S' of the spherical substrate. Note that, in the crater model, this contact line is depicted by a circle that is shrinking towards a singular point on a flat surface, as shown in Figs. 2(c) and 2(e).

Our analysis in this study focuses on the '*complete wetting*' regime as well, in which the equilibrium contact angle of the fluid reaches to zero. The available experimental results in this paradigm as well as reviews of this subject suggest that the contact line is preceded by a microscopic precursor film; therefore, one can neglect the friction at the contact line and consider in the precursor film instead.[10] However, in the present analysis, we consider the contribution of the line tension at the contact line in the viscous dissipation to study the spreading phenomenon.[5]

In an effort to capture the flow physics of our interest, including the phenomenon during the late-stage spreading of a viscoelastic drop, we now solve the Cauchy momentum equation which is given by [14,38]

$$\rho\left[\frac{\partial u}{\partial t} + (u.\nabla)u\right] = -\nabla p + \nabla.\boldsymbol{\tau} \qquad (2)$$

where $p$ is the pressure, $\rho$ is the density and $\boldsymbol{\tau}$ is the shear stress. In the present study, we assume that the flow of fluid in the drop volume is steady and mildly influenced by the inertia, i.e. the flow has a low Reynolds number $(\text{Re} \ll 1)$. Accordingly, the Cauchy momentum equation can be reduced to:[14]

$$-\nabla p + \nabla \cdot \boldsymbol{\tau} = 0 \qquad (3)$$

It may be mentioned in the context of Eq. (3) that we neglect the effect of gravitational force in the underlying spreading dynamics as the capillary length of the drop is much higher than that of the radius of the drop.[39] For the clarification on this part, one may refer to the corresponding order of magnitude analysis as given later in section III of this paper. We use the following boundary conditions to solve Eq. (3) as mentioned below.

$$\left.\begin{array}{l}\text{No shear at the liquid-air interface}: \partial u/\partial \mathbf{n} = 0 \\ \text{No slip at the solid-liquid interface}: \mathbf{n} \times u = 0\end{array}\right\} \qquad (4)$$

where $\mathbf{n}$ is the normal on a drop surface. We now look at Eq. (1) and make an effort to obtain the expression of shear stress, which is required to solve for the velocity field of a viscoelastic liquid. It is worth mentioning here that our analysis of drop spreading, consistent with the energy balance approach, largely relies on the accurate prediction of dissipative energy due to friction at the bulk volume and the line tension at the contact line. Note that in the present study, we do not consider the friction or shear resistance at the contact line to calculate the viscous dissipation therein since this aspect demands for a model consistent with the molecular kinetic theory (MKT).[5,10] An accurate prediction of the total dissipative energy necessitates the description of liquid velocity in the flow field. Next, we discuss the intermediate steps for the derivation of stress using Eq. (1), essentially to obtain the velocity field in the drop volume. Note that this exercise will help readers to have a complete understanding of the underlying method adopted in this study. To start with, for the problem under the present investigation, we consider the cylindrical coordinate system with axial symmetry; since the advancement of the crater towards the singular point 'S' (South Pole) will be axially symmetric (i.e., $\partial()/\partial\theta = 0$ and $u_\theta = 0$). Now, using Eq. (1), we get,

$$\boldsymbol{\tau} = \begin{pmatrix} \tau_{rr} & \tau_{rz} \\ \tau_{zr} & \tau_{zz} \end{pmatrix} \text{ and } u = \begin{pmatrix} u_r \\ u_z \end{pmatrix} \tag{5}$$

The deformation rate tensor takes the following form as:

$$\mathbf{D} = \frac{\left(\nabla u + (\nabla u)^T\right)}{2} = \begin{pmatrix} \dfrac{\partial u_r}{\partial r} & \dfrac{1}{2}\left(\dfrac{\partial u_r}{\partial z} + \dfrac{\partial u_z}{\partial r}\right) \\ \dfrac{1}{2}\left(\dfrac{\partial u_r}{\partial z} + \dfrac{\partial u_z}{\partial r}\right) & \dfrac{\partial u_z}{\partial z} \end{pmatrix} \tag{6}$$

Assuming the flow to be in the creeping flow regime (low Reynolds number flow), which is valid for present analysis as well, and $u_z \approx 0$, we can write the following equations for the problem under consideration as:

$$\left. \begin{array}{l} \dfrac{\partial u_r}{\partial r} = 0 \\[6pt] \dfrac{\partial \tau_{rr}}{\partial r} = \dfrac{\partial \tau_{rz}}{\partial r} = \dfrac{\partial \tau_{zz}}{\partial r} = 0 \end{array} \right\} \tag{7}$$

Now, employing this condition as given in Eq. (7) and upon substituting the expression of $\boldsymbol{\tau}$, $u$ and $\mathbf{D}$ [Eqs. (5) and (6)], the constitutive equation for the viscoelastic fluid, as given by Eq. (1), takes the following form:

$$f(\text{tr}(\boldsymbol{\tau})) \begin{pmatrix} \tau_{rr} & \tau_{rz} \\ \tau_{zr} & \tau_{zz} \end{pmatrix} - \lambda \begin{pmatrix} 2\tau_{rz}\dfrac{\partial u_r}{\partial z} & \tau_{zz}\dfrac{\partial u_r}{\partial z} \\ \tau_{zz}\dfrac{\partial u_r}{\partial z} & 0 \end{pmatrix} = 2\eta \begin{pmatrix} 0 & \dfrac{1}{2}\left(\dfrac{\partial u_r}{\partial z}\right) \\ \dfrac{1}{2}\left(\dfrac{\partial u_r}{\partial z}\right) & 0 \end{pmatrix} \tag{8}$$

From Eq. (8), we obtain the following set of equations as written below.

$$f(\text{tr}(\boldsymbol{\tau}))\tau_{rr} = 2\lambda \tau_{rz}\frac{\partial u_r}{\partial z} \tag{9.a}$$

$$f(\text{tr}(\boldsymbol{\tau}))\tau_{rz} = \eta\frac{\partial u_r}{\partial z} + \lambda \tau_{zz}\frac{\partial u_r}{\partial z} \tag{9.b}$$

$$f(\text{tr}(\boldsymbol{\tau}))\tau_{zz} = 0 \tag{9.c}$$

It is important to mention here that, for the assumption considered in this analysis, Eq. (9) as delineated above describe the rheology of the viscoelastic fluid consistent with the sPTT model. From Eq. (9.c), we get $\tau_{zz}=0$, since $f(\text{tr}(\boldsymbol{\tau}))=0$ will give a trivial solution. Now substituting $\tau_{zz}=0$ [from Eq. (9.c)] in Eq. (9.b), we get the following equation.

$$f(\text{tr}(\boldsymbol{\tau}))\tau_{rz} = \eta \frac{\partial u_r}{\partial z} \tag{10}$$

The function of the trace of the stress tensor is ideally an exponential function. However, for a very low shear stress magnitude, this function can be approximated using its linear form as given below.[14,24]

$$f(\text{tr}(\boldsymbol{\tau})) = 1 + \frac{\varepsilon\lambda}{\eta}\tau_{rr} \tag{11}$$

where $\varepsilon$ is the extensibility parameter used to model the extensional viscosity of the fluid; instead, this parameter, to be precise, limits the extensional viscosity, and $\lambda$ is the relaxation parameter. Now, we look at Eqs. (9.a) and (10) to obtain the following,

$$\tau_{rr} = \frac{2\lambda}{\eta}\tau_{rz}^2 \tag{12}$$

Using Eqs. (10), (11) and (12), we obtain another form of the governing equation, as written below.

$$\tau_{rz}\left(1 + \frac{2\varepsilon\lambda^2}{\eta^2}\tau_{rz}^2\right) = \eta\frac{\partial u_r}{\partial z} \tag{13}$$

The above equation [Eq. (13)] is the final form of the governing equation for describing the flow dynamics of the viscoelastic fluid pertinent to this analysis.[14,24] Using this constitutive relation, we solve the Cauchy momentum equation as described in Eq. (3) essentially to obtain the velocity distribution in the drop volume. We next write the Cauchy momentum equation for low Reynolds number and axially-symmetric flow as:

$$-\frac{dp}{dr} + \frac{\partial \tau_{rz}}{\partial z} = 0 \tag{14}$$

Now using this equation [Eq. (14)], we calculate the order of magnitude of the shear stress $\tau_{rz}$ by appealing to the condition of zero stress condition at the free surface, i.e., $\tau_{rz}|_{z=h(r)}=0$. The expression of shear stress reads as:

$$\tau_{rz} = \frac{dp}{dr}(z-h) \tag{15}$$

where $h(r)$ is the height of the liquid-air surface from the substrate at the distance $r$ from the drop center [Fig. 2(c)]. Using Eqs. (13), (15) and no-slip (dynamic boundary condition) condition at the surface, i.e. $u_r|_{z=0} = 0$ we get,

$$u_r = \frac{1}{\eta}\frac{dp}{dr}\left[\left(\frac{1}{2}\left\{(z-h)^2 - h^2\right\}\right) + \left(\frac{2\varepsilon\lambda^2}{4\eta^2}\right)\left(\frac{dp}{dr}\right)^2\left\{(z-h)^4 - h^4\right\}\right] \qquad (16)$$

Note that for $\lambda = 0$, the velocity distribution described in Eq. (16) confirms the velocity field for a Newtonian fluid on spherical substrates.[5] It may be mentioned here that Eq. (16) governs the relation among the flow parameters, i.e., the fluid velocity $(u_r)$ and the fluid properties, i.e., the viscoelastic parameters $\lambda$, and the extensibility parameter $\varepsilon$. On a related note, we would like to mention here that the aforementioned viscoelastic parameters can be grouped together and represented by a non-dimensional parameter. This non-dimensional parameter is termed as the Deborah number or Weissenberg number.[24] Also, it is important to mention in the context of Eq. (16) that the pressure gradient $dp/dr$ is a crucial parameter for the present system, and its effect can be determined using the order of magnitude of analysis. To perform such analysis, we use Eq. (14) and compare the scales of pressure gradient $dp/dr$ and the gradient of shear stress $d\tau_{rz}/dz$. The basis of this comparison suggests that the spatial changes in the shear stress $\tau_{rz}$ can be accounted for the variation in the pressure gradient $dp/dr$ generated during the spreading phenomenon.[16] The scale of $\tau_{rz}$ is proportional to $(\eta u_{ref})/h$ (refer to Eq. (13)). On using this scale, we estimate the order of pressure gradient $dp/dr$ as:

$$\frac{dp}{dr} \sim \frac{\eta u_{ref}}{h^2} \qquad (17)$$

where $u_{ref}$ is the late-stage spreading (precisely, the contact line velocity) velocity which can be estimated by knowing the substrate radius $R$ and the instantaneous magnitude of the angle $\phi$. The pressure gradient will have a negative value, which implies that the proportionality constant has to be negative. Moreover, since we are equating the order of the magnitude of the terms in Eq. (14), this proportionality constant will be unity. Important to mention, this proportionality constant will assume different values for the varying magnitude of the fluid viscoelasticity. Note that with a change in the relaxation time of the fluid, the viscoelasticity of the fluid changes as realised by the different values of the Deborah number in the dimensionless form.

Now, Eq. (16) can be written in the following dimensionless form as:

$$\bar{u}_r = \frac{-1}{2}\left[\left\{1-(1-\bar{z})^2\right\}\left\{1+\varepsilon De^2\left(1+(1-\bar{z})^2\right)\right\}\right] \tag{18}$$

where $De\ (=\lambda u_{ref}/h)$ is the Deborah number, $\bar{u}_r\ (=u_r/u_{ref})$ is the dimensionless velocity and $\bar{z}\ (=z/h)$ is dimensionless coordinate. As mentioned before, the Deborah number is a dimensionless number signifying the elastic effect of the fluid.[24] It is worth mentioning here that the expression of velocity distribution as given in Eq. (18) for a limiting case of $De=0$ matches well with the expression reported by Iwamatsu for Newtonian fluids $(n=1)$.[5] Since we consider the energy balance approach in this study to obtain the drop spreading behaviour, we now calculate the viscous dissipation in the fluid drop. The expression of the viscous dissipation $\Phi$ can be written as:

$$\Phi = \int_\Omega \varphi\, d\Omega \tag{19}$$

Here, $\varphi\ (=\tau_{rz}(du_r/dz))$ is the viscous dissipation per unit volume of the drop $d\Omega$ and $\Omega$ is the total volume.[14,40] For axisymmetric flow, the viscous dissipation can be written as:[14]

$$\Phi = \int_{r_L+\Delta r}^{r_M}\int_0^{2\pi}\int_0^h \tau_{rz}\left(\frac{du_r}{dz}\right)dz\, d\phi\, rdr \tag{20}$$

It is worth mentioning here that to avoid the problem of singularity at $r=0$, we here introduce the upper bound $r_M$ and the lower bound $r_L$ with the cut-off radius $\Delta r$. Note that Eq. (20) can be integrated to obtain the explicit expression of viscous dissipation. We can calculate $\tau_{rz}$ from Eqs. (15) and the velocity gradient from Eq. (18). Below, we write the expression of velocity gradient $(du_r/dz)$, which can be obtained from Eq. (18) as:

$$\frac{d\bar{u}_r}{d\bar{z}} = -\left[(1-\bar{z})\left\{1+2\varepsilon De^2(1-\bar{z})^2\right\}\right] \tag{21}$$

Now, substituting the expressions of $\tau_{rz}$ and $(du_r/dz)$ in Eq. (20), and then integrating this equation [Eq. (20)] with respect to $z$ and $\phi$, we will get the expression of the viscous dissipation. The form of viscous dissipation reads as:

$$\Phi = 2\pi\eta u_{ref}^2\left[\frac{1}{3}+\frac{2\varepsilon De^2}{5}\right]\int_{r_L+\Delta r}^{r_M}\frac{1}{h}rdr \tag{22}$$

Since we have considered a crater model in this analysis, we can assume the meniscus of the viscoelastic drop, as shown by Figs. 2(c) and 2(e) to be axially-symmetric wedged-shaped.

Now, owing to this consideration, we can write the following expression for a small contact angle $\theta$ as:

$$h(r) = \tan\theta(r - r_L) \sim \theta(r - r_L) \qquad \forall \text{ small } \theta \tag{23}$$

Further integration of Eq. (22) yields the following final expression of the viscous dissipation and given as:

$$\Phi = \frac{2\pi\eta u_{ref}^2 r_L \kappa}{\theta}\left(\frac{1}{3} + \frac{2\varepsilon De^2}{5}\right) \tag{24}$$

The term $\kappa$ appearing in Eq. (24) can be written below:

$$\kappa = \Lambda - 1 + \ln(\Lambda - 1) - \delta - \ln\delta \tag{25}$$

where $\Lambda = r_M/r_L$ and $\delta = \Delta r/r_L$. Note that Eq. (25) shows a well-known singularity, which is $\kappa \sim \ln\delta$. Important to mention here, this singularity will occur as $\delta \to 0$. But for the present problem of spreading on a spherical substrate where $r_L \to 0$, this singularity $(\delta \to 0)$ will not be important, and hence, it will not occur. Next, we calculate the thermodynamic force (capillary force), which also includes the free energy at the three-phase (fluid-fluid-solid) contact line. Below we write the expression of the thermodynamics force (capillary force per unit length) as:[6]

$$f_{cap} = \sigma_{lv}(\cos\theta_Y - \cos\theta) - \frac{\tilde{\lambda}}{R\tan\phi} \tag{26}$$

In Eq. (26), $\sigma_{lv}$ is the liquid-air surface tension, $\tilde{\lambda}$ is the line tension at the three-phase contact line,[6] $\theta_Y$ is Young's contact angle at equilibrium, and $\phi$ is half of the central angle, as shown in Fig. 2(a). Now, the energy balance condition at the contact line with the radius $r_L$ leads to the following equation as written below:[14]

$$2\pi r_L f_{cap} u_{avg} = \Phi \tag{27}$$

Note that $u_{ref} \sim u_{avg}$. Now using Eqs. (20)-(21) and (23)-(25), we can rewrite Eq. (27) in the form as given below:

$$\sigma_{lv}\left[(\cos\theta_Y - \cos\theta) - \frac{\tilde{\lambda}}{\tan\phi}\right] = \frac{\eta u_{ref}\kappa}{\theta}\left(\frac{1}{3} + \frac{2\varepsilon De^2}{5}\right) \tag{28}$$

In Eq. (28), $\tilde{\lambda} = \lambda/(\sigma_{LV}R)$ is the scaled line tension relative to liquid-air interfacial tension $\sigma_{lv}$.[4–6] As already mentioned, our analysis takes into account the phenomenon during the late-stage spreading of an elastic non-Newtonian drop on a spherical substrate. We here make an effort to investigate the variation of the contact angle on a hydrophilic surface and its effect on the variation of contact-line velocity, as discussed next. For late-stage spreading on a hydrophilic substrate, we will consider $\theta \ll 1$ and $\theta_Y \ll 1$, while the angle $\psi$ following its description in Fig. 2(a) can be written as:

$$\left.\begin{array}{l} \psi\big|_{\text{late-stage}} = (\pi - \phi) \to 0 \\ r\big|_{\text{late-stage}} \to r_0 \end{array}\right\} \tag{29}$$

Note that $r_0$ in the above equation is the radius of the drop as it completely wets the spherical substrate of radius R, i.e., $r_0 > R$. Hence, in a complete wetting condition $(i.e., \theta_Y = 0)$, we can obtain the drop volume $V_0 = (4\pi/3)(r_0 - R)^3$. Also, having a closer look at Fig. 1, we get the following expression as written below.

$$\tan \phi = \frac{r \sin \theta}{R - r \cos \theta} \tag{30}$$

Now using Eq. (29) and including the consideration leading to the late-stage spreading behaviour of the drop, we obtain the following expression as:

$$\tan \phi \approx -\psi \to \frac{-r_0}{(r_0 - R)} \theta \tag{31}$$

Hence, the radius of the contact line $(r_L)$ can be written as (cf. Fig 1 for a clearer view):

$$r_L = R \sin \psi \simeq R\psi \simeq \frac{r_0}{(r_0 - R)} R\theta \tag{32}$$

Now using Eqs. (29)-(32), the equation governing the spreading dynamics (energy balance condition), i.e., Eq. (28), which includes the late-stage scenarios as well, takes the following form.

$$\theta \left[ \frac{1}{2}(\theta^2 - \theta_Y^2) + \frac{r_0 - R}{r_0 \theta} \tilde{\lambda} \right] = \frac{\eta u_{ref} \kappa}{\sigma_{LV}} \left( \frac{1}{3} + \frac{2\varepsilon De^2}{5} \right) \tag{33}$$

As the contact angle $\theta$ during late-stage spreading is very low, the spreading velocity (precisely, the contact line velocity) on the spherical substrate can be written as:

$$u_{ref} \sim \frac{d}{dt}(R\phi) = \frac{d}{dt}(-R\psi) = -\frac{r_0}{(r_0 - R)} R\dot{\theta} \tag{34}$$

Now, substituting $u_{ref}$ in Eq. (33), we get another form as written below.

$$\theta \left[ \frac{1}{2}(\theta^2 - \theta_Y^2) + \frac{r_0 - R}{r_0 \theta} \tilde{\lambda} \right] = -\frac{\kappa \eta}{\sigma_{LV}} \left( \frac{1}{3} + \frac{2\varepsilon De^2}{5} \right) \frac{r_0}{r_0 - R} R\dot{\theta} \tag{35}$$

We proceed one step further to cast Eq. (35) into the following form as given by,

$$\Gamma\left(\frac{1}{3}+\frac{2\varepsilon De^2}{5}\right)\frac{d\theta}{dt}+\theta\left[\frac{1}{2}\left(\theta^2-\theta_Y^2\right)+\frac{r_0-R}{r_0\theta}\tilde{\lambda}\right]=0 \qquad (36)$$

The coefficient $\Gamma=\frac{\kappa\eta}{\sigma_{lv}}\left(\frac{r_0}{r_0-R}\right)R$ is the time scale of spreading. As clearly evident from the expression, this coefficient depends on the radius of the substrate $R$, and the drop volume $V_0$ having radius $r_0$ $\left(i.e., V_0=(4\pi/3)(r_0-R)^3\right)$. It needs to be mentioned here that, for $De=0$, Eq. (36) reduces to the equation, governing the spreading dynamics of the Newtonian drop on the spherical substrate.[4,6]

$$\frac{d\theta}{dt}=-\frac{\theta}{\Gamma}\left[\frac{1}{2}\left(\theta^2-\theta_Y^2\right)+\frac{r_0-R}{r_0\theta}\tilde{\lambda}\right] \qquad (37)$$

Now for the complete wetting case, i.e., for an underlying spreading phenomenon on a completely hydrophilic surface, as realisable by Young's contact angle $\theta_Y=0$, and by neglecting the line tension effect ($\tilde{\lambda}=0$), we can write Eq. (37) in the following form.

$$\theta=\theta_0\left(1+\theta_0^2\left(t/\Gamma\right)\right)^{-1/2} \qquad (38)$$

where $\theta_0$ is the initial contact angle at $t=0$. Now we can write the time evolution of dynamic contact angle as:

$$\theta\propto\left(t/\Gamma\right)^{-1/2} \qquad (39)$$

It is worth mentioning here that Eq. (37) as obtained for the limiting condition of $De\to 0$ is showing similarity with the reported equation governing the spreading of Newtonian drop $(n=1)$ on the spherical substrate.[5,6] We would like to mention here that for the smaller value of dynamic contact angle $(\theta\to 0)$, the contributory effect of line tension [cf. Eq. (35); the second term of LHS of this equation becomes effective as $\theta\to 0$] on the underlying spreading dynamics becomes important. Accounting this aspect, we solve Eq. (36) using the Runge-Kutta method for an initial contact angle $\theta_0\sim 1$. Also, we consider ring spreading on a complete hydrophilic substrate $(\theta_Y=0)$, and the effect of line tension is taken comparatively in a smaller range.

## B. Experimental details

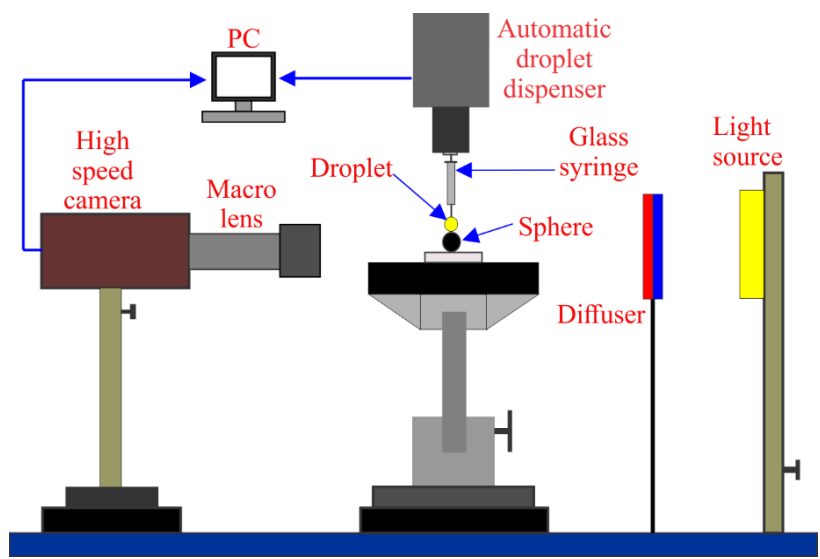

**Figure 3 (will appear colour online**): Schematic of the experimental set-up. The set-up includes the following: Macro lens, high-speed camera, automatic drop dispenser, spherical substrate placed on the test plane, diffuser, light source, glass syringe.

To verify the accuracy of the theoretical model outlined in the preceding section, we perform experiments on the spreading of a viscoelastic drop on a spherical substrate. In particular, the experimental observations substantiate the credibility of our theoretical model in describing the phenomenon during the late-stage of spreading as well. We show, in Fig. 3, the experimental set-up schematically. Drops of a predefined size $(V_0 = 3\mu l)$ are generated using a 250 µl glass syringe (Make: Hamilton), which is further controlled by a precision drop dispenser mechanism (Make: Apex instruments), integrated with the circuit. We capture the spreading dynamics at 50 frames/s (1280 × 800 pixels resolution) by using a high-speed camera (Make: Phantom), coupled with a macro lens (Make: Nikkor, Nikon) of focal length 105 mm. For the optical arrangement, as assimilated in the circuit, we use a brightness controlled white LED light (Make: Holmarc Opto-Mechatronics) along with a diffuser as a backlight. Notably, for the present experiments, we use the solution of 5% Polyisobutylene (PIB) in Tetradecane (C14) as the viscoelastic fluid. A polished stainless-steel ball of 3.85mm diameter is used as the spherical substrate, as shown in the schematic of Fig. 3. The cleaning protocol is such that the substrate is initially cleaned with acetone and then dried in a hot air oven at $120^0$C for 20 minutes. Note that all the experiments are conducted in the capillary controlled spreading regimes. Also, being an integral part of the present endeavour, we perform the spreading experiments of a Newtonian drop (De-ionised water) as well. To ensure that the Newtonian droplet exhibits complete wetting on the solid substrate, a surfactant solution (0.01% volume fraction of lauric acid) is added to it. Note that, in the present experiments, we have used two sets of cameras to ensure that symmetricity in the droplet spreading phenomena is maintained. A high-speed camera and a CMOS camera (placed perpendicular to the direction of the high-speed camera) are employed for verifying the alignment between the glass syringe needle and the spherical substrate. This alignment ensures a proper discharge of the droplet (from the syringe needle) onto the centre of the sphere.

### III. Fluid preparation

Here we briefly discuss the fluid preparation of the viscoelastic fluid used in this study. For the preparation of the viscoelastic fluid, we chemically synthesise a solution of Polyisobutylene (PIB-Vistanex L120, $M_w = 1 \times 10^6$) in Tetradecane (C14-$C_{14}H_{30}$-Newtonian solvent). The characterisation of the PIB/C14 solution is found in the referred studies.[27,31] This fluid is prepared as follows: small pieces of PIB (5%, w/w) were thoroughly mixed in tetradecane (solvent) with the help of a magnetic stirrer (Make: Tarsons). The approximate duration of the stirring process was around five days. Following this, the prepared solution was continuously rotated on a rolling machine (Make: LMBR500) for one week. It is important to mention that the ambient temperature throughout the duration of the fluid preparation was maintained around 25°C. For this particular volume fraction (5%, w/w) of PIB in PIB/C14 solution, the rheology of the prepared viscoelastic fluid, as reported in seminal works, deems to be fitted best by the Phan-Thien-Tanner (PTT) model.[27,31,32,41,42] Therefore, the parameters such as extensibility parameter, dynamic viscosity and the relaxation time can be obtained from the referred studies. The present study does not necessitate the separate estimation of these parameters. Accordingly, following this literature[27,31,32,41,42], we use in the present analysis: viscosity $\eta = 1.424$ Pa.s, relaxation time $\lambda = 0.06$ sec, and extensibility parameter $\varepsilon = 0.05$, for PIB/C14 solution. The surface tension and the scaled line tension of the prepared fluid were found to be around 22 mN/m and 0.02, respectively.[6,27] Now, using all the values mentioned above, we calculate the reference scales, i.e., the velocity scale $u_{ref}$ and the length scale $h$ to be $u_{ref} \sim 10^{-3}$ m/sec, and $h \sim 10^{-5}$ m, respectively. It is worth mentioning here that for these reference scales and using the parameters mentioned above; we calculate the magnitude of the Capillary number $Ca$ and the viscoelastic parameter $\varepsilon De^2$ to be of the order of $O\left[10^{-3}\right]$ and $O\left[10^{-1}-10^{0}\right]$.

It is important to mention here that the literature on the use of sPTT or PTT model to depict the rheology of (any) Boger fluid is very sparse and therefore, thus the spreading analysis of Boger fluid due to insufficient data remains beyond the scope of the present work. Note that the solution of 5% PIB in C14 (solution of Polyisobutylene in Tetradecane) represents a shear-thinning viscoelastic fluid.[43] From the physical properties of the viscoelastic fluid considered in this analysis, i.e. the solution of PIB/C14, and the drop volume of 3 $\mu l$, it appears that the capillary length of the drop is much higher than the radius of its representative spherical volume. Capillary length is the characteristics length scale which helps to decide the effect of gravitational force on the spreading drop. If it is higher than the radius of the drop, then the effect of gravity can be neglected. In the present case, the capillary length $\left(=\sqrt{\sigma_{lv}/(\rho_{PIB/C14}g)}\right)$ is 1.6743 mm, and it is greater than that of the radius of the drop, which is 0.894 mm (according to 3 $\mu l$). Therefore, in the present study, we neglect the effect of gravitational force. This assumption is previously mentioned in the context of Eq. (3).

## IV. Results and Discussion

### A. Experimental observations of the spreading: Theoretical Model Benchmarking

As an important objective of the present endeavour, we here make an effort to establish the effectiveness of our theoretical model developed in predicting the drop spreading behaviour on a spherical substrate. In doing so, results obtained from the present theoretical modelling framework are compared with our experimental observations. We show in Fig. 5(a) the temporal variation of $d\bar{h}/d\bar{t}$ during the spreading process, while the snapshots of the drop spreading process on the spherical substrate captured at various time instants are shown in Fig. 5(b)-(c). The experimental results are obtained for both the Newtonian (Fig. 5(c)) and viscoelastic fluids (Fig. 5(b)). The markers in Fig. 5(a) are used to represent the experimental results, while the straight line shows the results obtained from the theoretical calculations. As previously mentioned, $\bar{h}$ is the normalized height (normalized by the initial height) or a distance between the top surface of the spreading droplet and that of the spherical substrate (or the surface of a spherical substrate on the North Pole (refer to Fig. 4)). The volume of the spreading drop in the present study is assumed to remain constant, and therefore, a cruising motion of the contact line is proportional to the reduction in the concerned height $\bar{h}$. This signifies the rate of change in the height $\bar{h}$ indirectly manifests as the variation in the contact line velocity for the present analysis (refer to Fig. 4). Otherwise, in experiments, it is a tedious work to track the location of the contact line due to microscopic scale of the droplet's precursor film being formed on the spherical substrate (refer to Fig. 5(b), particularly a late-stage spreading).

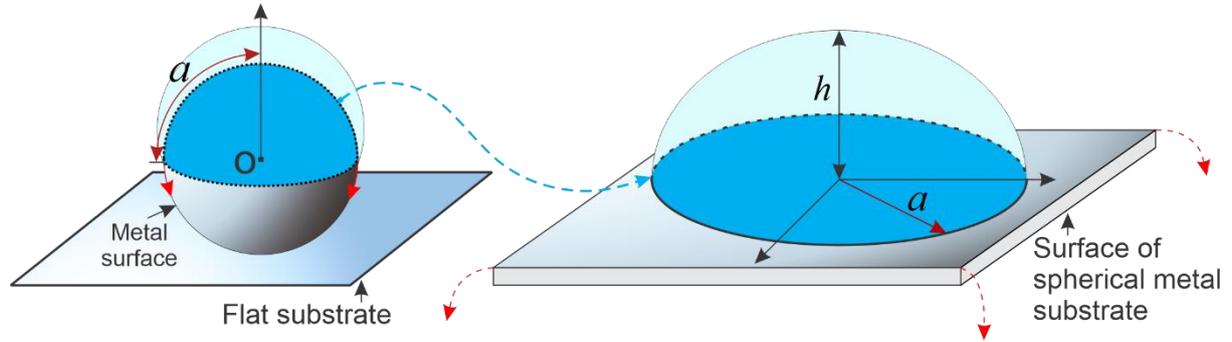

**Figure 4 (color online):** Schematic showing the equivalence of the hemispherical droplet's footprint at a particular instance of spreading on the spherical substrate with the corresponding footprint on the flat substrate. The peripheral distance between a point on the spherical substrate lying on the vertical axis and the point on the contact line of the droplet, i.e. $a$, is equivalent to the radius of the droplet's footprint on the flat substrate. Here, $h$ is the height of the droplet's open boundary from its footprint.

It may be reiterated here that the present study focuses on the complete wetting phenomenon which incorporates such precursor films, and ideal and real contact lines.[10] The relation between the height $\bar{h}$ and the contact line location '$a$' on the spherical substrate can be obtained by looking at the volume of spreading droplet [Eqs. (40) and (41)]. Location $a$ refers to the distance of the contact line from the initial contact of the liquid droplet with the spherical substrate. For the simplification in the understanding, we here consider the volume of drop that is being spread on the flat substrate (refer to Fig. 4 for schematic depiction and Eq. (40) for mathematical expression). It can be mentioned here that this assumption is well applicable for the present study since the volume of the droplet (3 $\mu l$) is very small as compared to the volume or size of the spherical substrate, i.e., 30 $\mu l$ considered for the present study. Moreover, the homogeneous as well as the quasistatic spreading of the droplet, as is the case in the present scenario, can be taken into account in the present analysis (refer to the temporal evolution of the spreading droplet shown in Fig. 5(b), particularly in the regime of late-stage spreading). Accordingly, as shown in Fig. 4, we can assume that the hemispherical footprint of the droplet at particular instance of spreading is similar to the equivalent droplet footprint on the flat substrate. For such case, the approximate volume of the droplet can be given as:

$$V_0 = \frac{\pi h}{6}\left(3a^2 + h^2\right) \qquad (40)$$

Note that height $h$ and distance $a$, both are a function of time. In Eq. (40), the parameter $a$ can be estimated by the relation,

$$\tan\theta = \frac{h}{a} \qquad (41)$$

On using this relation in Eq. (40), the height $h$ can be estimated as,

$$h(t) = \left[\frac{\pi}{6V_0}\left(\frac{3}{\tan^2\theta} + 1\right)\right]^{-1/3} \qquad (42)$$

Note that $h(t)$ is an instantaneous height which varies during the spreading phenomenon with the instantaneous contact angle $\theta$. The plots shown in Fig. 5(a) are obtained using this relation [Eq. (42)].

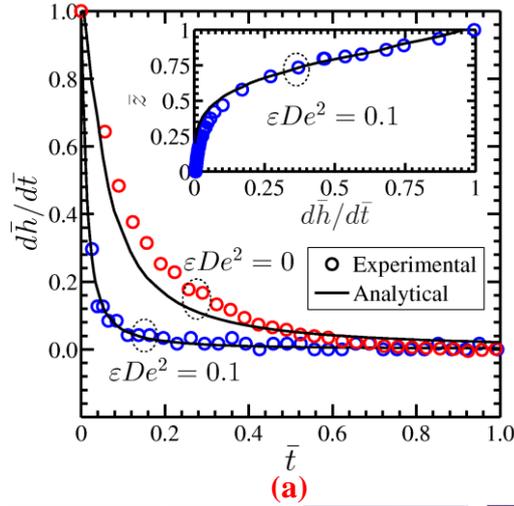

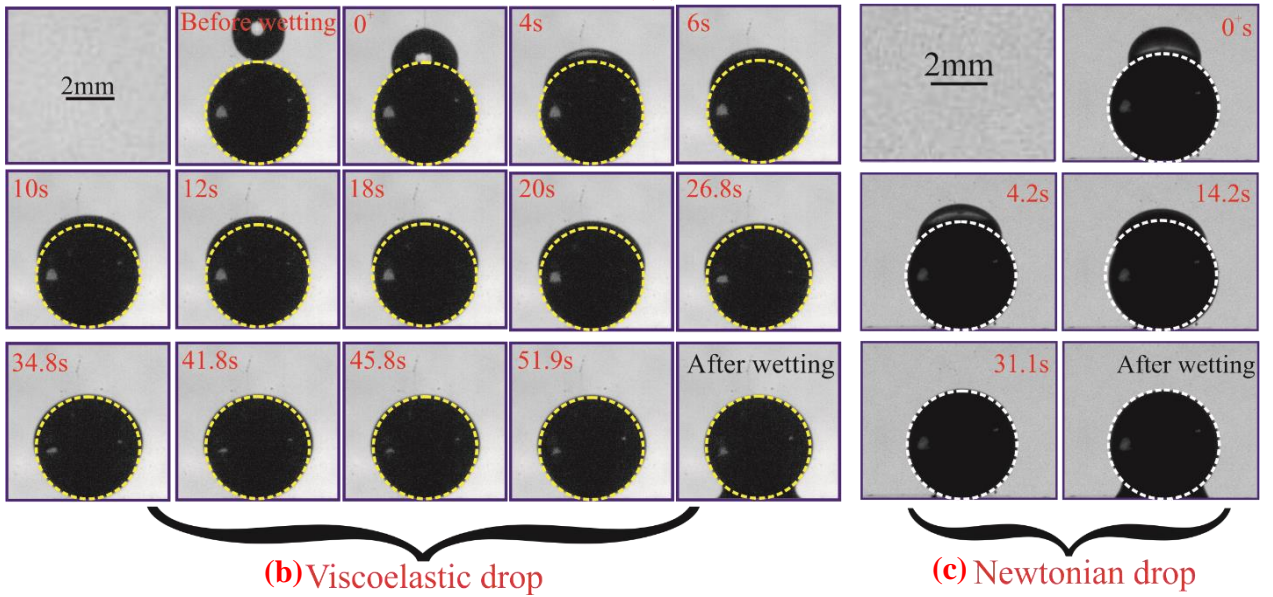

(b) Viscoelastic drop

(c) Newtonian drop

**Figure 5 (will appear color online**): (a) Plot shows the experimental and theoretical variation of $d\bar{h}/d\bar{t}$ along the temporal direction for both the viscoelastic and Newtonian fluids. The inset shows the variation of $d\bar{h}/d\bar{t}$ along $\bar{z}$ during the drop spreading process on the spherical substrate for viscoelastic fluid. (b) and (c) Snapshots show the temporal evolution of the drop spreading phenomenon of the viscoelastic fluid and Newtonian fluid respectively, on the spherical substrate. The dotted yellow circle demarcates the boundaries of the spherical substrate.

For these validation plots depicted in Fig. 5(a), we consider the spreading of a viscoelastic drop having viscoelasticity as represented by $\varepsilon De^2 \approx 0.1$. For the present study, Deborah number $(De)$ is 1.414 as the extensibility parameter for PIB/C14 solution is 0.05.[27] For the prevailing situation of the complete wetting phenomenon as considered here, it is indeed tedious work to calculate the shear rate at the contact line due to the microscale thickness of the precursor film which recede the contact line being formed. Therefore, in the present analysis, we have invoked to the scale of the shear rate instead of the actual shear rate in the analysis. The scale of shear rate can be estimated using the order of reference velocity and that of the thickness of the film (or completely spread droplet) which is of the order of a few microns.

We now discuss the notable points observed form the temporal evolution of $d\bar{h}/d\bar{t}$ in Fig. 5(a) as follows. During the initial stages of spreading, the temporal gradient is high, and we observe a sharp change in the temporal evolution of the height. We would like to mention here that the three-phase contact line plays an important role in governing the flow dynamics in the capillary-driven spreading phenomenon. In this initial regime of the spreading, the capillary force governs the flow by dominating its effect over the viscous forcing and results in a higher velocity of the contact line. It is because of the higher contact line velocity; we observe a sharp gradient of $d\bar{h}/d\bar{t}$ during the initial stage. Notably, this observation is valid for both the Newtonian and viscoelastic fluids. We observe different scenarios at a later stage of the spreading. Note that, as the time grows (later stage), the drop spreads on the substrate following a balance between the capillarity and the viscous stresses. Prominently, compatible with a capillary-viscous force balance, the contact line velocity slowed down during the late-stage of the spreading as supported by the snapshots presented in Fig. 5 (b)-(c) for both the Newtonian and viscoelastic fluids. Precisely, slower movement of the contact line velocity results in an insignificant change in the height with time, as witnessed by the constant value of $d\bar{h}/d\bar{t}$ in Fig. 5(a) and the images in Fig. 5(b) and 5(c). The inset in Fig. 5(a) plots the variation of $d\bar{h}/d\bar{t}$, during the late stage of the spreading process. Quite remarkably, the calculations predicted for both the Newtonian and viscoelastic fluids, by the present theoretical framework are in coherence with our observations made from our experimental investigations, as witnessed by Fig. 5(a). This observation justifies the credibility of our theoretical model developed in this study in predicting the spreading dynamics of an elastic non-Newtonian fluid.

In the context of the variation in Figs. 5(a)-(c), we observe another important aspect. In Fig. 5(a), we observe that the Newtonian fluid has a higher $d\bar{h}/d\bar{t}$, and hence takes less time to achieve complete wetting in contrast to the same for the viscoelastic fluid as witnessed in Figs. 5(b) and 5(c). Viscoelastic fluid takes 51.9 secs, whereas the Newtonian fluid takes 31.1 secs. This typical variation in the spreading rate can be attributed to the effect of rheology modulated viscous dissipation in the representative volume of the drop. In the case of a Newtonian fluid, lesser magnitude of the velocity gradients in the drop volume as observed later in Fig. 6 leads to lower viscous dissipation as compared to the viscoelastic fluid. In effect, the surface tension becomes dominant and leads to higher spreading rate for the Newtonian fluid. The viscous dissipation in the viscoelastic fluid is higher due to the higher shear-thinning and elastic nature of the fluid. Note that this typical variation in the spreading rate, as seen in Fig. 5(a), also justifies the variation in the dynamic contact angle, which is depicted later in Fig. 7(a).

**B. Effect of fluid viscoelasticity on the spreading dynamics**

**1.** *The variation of the flow velocity: Contact line dynamics*

Before going for the discussion of the pertinent results as presented in this section, it may be mentioned here that increasing the magnitude of the relaxation time ($\lambda$) (equivalently for the higher Deborah number in the dimensionless form) enhances the shear-thinning and elastic nature of the fluid. The rheology of the viscoelastic fluid for a particular range of applied shear rate shows a shear-thinning nature, whereas, beyond this range, it represents the Newtonian fluid. In particular, at nearly no-shear rate condition, the viscoelastic fluid shows a Newtonian behaviour, and identified by its zero-shear viscosity.[44] With an increment in the shear rate further, the viscoelastic fluid due to multiple degrees of freedom sets internal extra normal stresses in the field and leads to elongation of its polymeric network strands. This phenomenon, as a result, causes a less resistance by the viscoelastic fluid to the applied shear rate, or in other words, a decrement in its own viscosity. Such a reduction in the viscosity also depends on the relaxation time or an extensibility parameter of the fluid or a viscoelastic parameter. Either an increment in the relaxation time or the extensibility parameter or both give rise to more elastic nature to the fluid, which in effect, promotes the generation of higher extra normal stresses in its volume. Consequently, the higher is the value of a viscoelastic parameter, higher will be the effect of extra normal stresses, and will result in higher shear-thinning nature.[44] Mathematically, this cause-effect relation can also be depicted by the final form of the constitutive equation of the sPTT model, i.e. Eq. (13). It can be observed from Eq. (13) that with an increment in either the shear stress or a viscoelastic parameter, the apparent viscosity of the fluid decreases as the zero-shear viscosity $\eta$ is divided by the term, which is always higher than 1 for a non-zero viscoelastic parameter, in the bracket (in LHS).

Because of this increasing elastic as well as shear-thinning nature, the fluid will behave more like a solid than a liquid for higher Deborah number.[45] Therefore, for the viscoelastic fluids having higher viscoelasticity, as realized by the higher value of $\varepsilon De^2$ (as with increasing the magnitude of relaxation time), the spreading ability will decrease. From Eq. (12), it is seen that the normal stress component ($\tau_{rr}$) and relaxation time ($\lambda$) are related by the expression $\tau_{rr} = (\tau_{rz})^2 2\lambda/\eta$. Following this expression, it is evident that $\tau_{rr}$ is directly proportional to the relaxation time of the fluid. As a consequence, the flow rate becomes higher for viscoelastic fluid as compared to that of a Newtonian fluid, while increasing relaxation time ($\lambda$) is expected to enhance the flow rate as well. We attribute this effect to the increase in the shear-thinning behaviour of the fluid with increasing viscoelasticity $(\varepsilon De^2)$.[24] Important to mention, the shear-thinning effect, which becomes stronger with increasing $(\varepsilon De^2)$, will increase the shear rate as can be seen in Eq. (18). From Eq. (18), we can get $du_r/d\bar{z} = -\left[(1-\bar{z})\left\{1+2\varepsilon De^2(1-\bar{z})^2\right\}\right]$. We now make an effort to represent this equation graphically in Fig. 6, wherein the non-dimensional velocity gradient is plotted with a change in drop height $\bar{z}$ (non-dimensional) for different values of $\varepsilon De^2$.

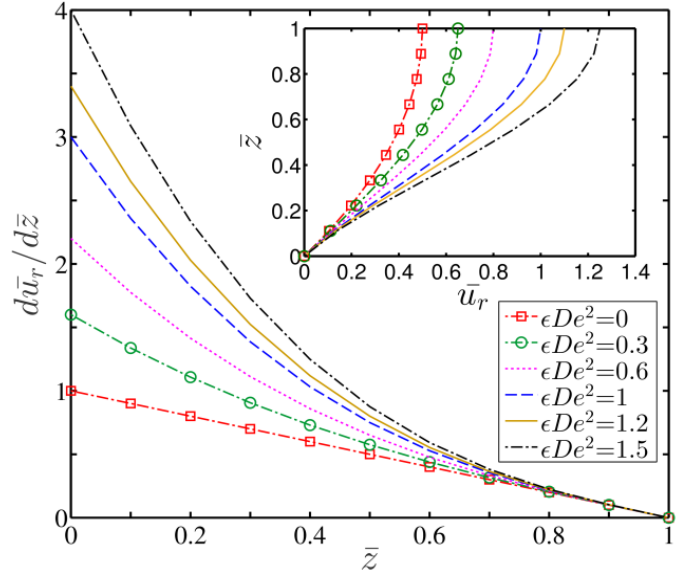

**Figure 6 (will appear colour online**): Plot showing the variation in velocity gradient (dimensionless) versus drop height, obtained for different values of $\varepsilon De^2$. As shown in the inset, with increasing viscoelasticity of the fluid as realised by the increasing value of $\varepsilon De^2$, fluid velocity in the drop volume becomes higher due to the higher shear-thinning effect of the fluid. The higher fluid velocity leads to an increment in the velocity gradient.

From Fig. 6, it is seen that the shear rate at the substrate $(\bar{z}=0)$ increases with the increasing the magnitude of the viscoelastic parameter. Also, one can find from Fig. 6 that the Newtonian fluid will have the least velocity gradient in the drop flow field, while the magnitude of the velocity gradient becomes higher for the larger viscoelasticity of the fluid. In the capillary driven regime, the higher flow velocity of the viscoelastic fluid, which increases with increasing the magnitude of $\varepsilon De^2$ as well (see inset of Fig. 6), leads to a higher shear rate in the drop volume. From the definition given in Eq. (18), viscous dissipation can be related as $\varphi = \tau_{rz}(du_r/dz)$, and hence, an increase in velocity gradient is expected to increase in the viscous dissipation. Since we are considering the energy balance approach in this study, which is consistent with the hydrodynamic model as well[2,14], the effect of viscous dissipation becomes the only opposing factor for the capillary driven spreading in the absence of friction (MKT model).[1,3,46] As mentioned before, an increase in the viscoelasticity of the fluid increases the viscous dissipation effect following enhancement in the magnitude of the velocity gradient therein. Notably, this augmented viscous dissipation effect leads to an increase in the magnitude of the dynamic contact angle (cf. Fig. 7(a) in the next subsection). As a consequence of this phenomenon, the time required for the complete wetting of the substrate becomes more for the higher values of $\varepsilon De^2$ that too is observed from Fig. 7(a). Important to mention here that the flow velocity in the drop volume, as shown in the inset of Fig. 6, can be approximated to the contact line velocity, since our analysis is focused on the capillary controlled spreading of the drop. Quite notably, the velocity in the drop flow field, as shown in the inset of Fig. 6, shows similar qualitative behaviour with our experimental observations, as shown in the inset of Fig. 5(a). This qualitative observation supports the efficacy of our theoretical model in predicting the drop spreading dynamics, including the phenomenon during late-stage. Although the spreading dynamics on the spherical substrate is expected to be different than that on a flat surface, the effect of the viscoelasticity will always try to reduce the spreading rate as discussed in the forthcoming section in greater detail.

**2. *The variation of the dynamic contact angle: A perspective of complete wetting***

We depict in Fig. 7(a) the variation in dynamic contact angle $\theta$ versus the non-dimensional time, obtained for different values of the viscoelastic parameter. Note that to demarcate the effect of viscoelasticity on the underlying spreading behaviour, we plot in Fig. 7(a) the variation obtained for the limiting case of $\varepsilon De^2 = 0$, i.e., for Newtonian fluid as well. From the variation portrayed in Fig. 7(a), it is seen that at any temporal instant, the dynamic contact angle of the viscoelastic drop is higher than that of a Newtonian liquid drop. Moreover, the dynamic contact angle increases with increasing the viscoelastic behaviour $\left(\varepsilon De^2\right)$ of the fluid, which is observed from Fig. 7(a) as well. Important to mention, we consider the capillary number to be of an order of $\text{Ca} \sim 10^{-3}$ in this analysis, implicating the underlying phenomenon of spreading to be compatible with capillary-viscous force balance. An increase in the viscoelastic parameter, as realised by an increase in solute concentration will increase the shear-thinning nature of the fluid. It is worth adding here that, for a given strength of the surface tension force (the capillary number is fixed), the fluid velocity will increase with increasing the shear-thinning nature of the fluid as the effective viscosity of the fluid decreases. We have discussed the variation in fluid velocity as well as its gradient in the drop flow field, obtained with a change in the viscoelastic parameter in the next sub-section. We mention here that because of this relatively higher shear-thinning nature of the fluid with increasing the magnitude of viscoelasticity, the dissipative effect gets strengthened. On account of this higher viscous dissipation, the spreading rate will decrease (also discussed in the context of Fig. 5(a)-(c)), and this phenomenon will result in an increment in the dynamic contact angle, as seen in Fig. 7(a). It is because of this decrement in spreading rate, the time required for completely wetting the spherical substrate will be larger for a higher value of $\varepsilon De^2$ as witnessed in Fig. 7(a).

Note that this variation in the wetting time with the viscoelastic parameter for the present study can also be delineated in terms of the following expression:

$$\frac{t}{\Gamma} \sim \left[\frac{1}{3} + \frac{2\varepsilon De^2}{5}\right]$$

This expression is obtained from the scaling analysis of Eq. (36). It can be observed from the above expression that the wetting time of the spreading drop depends on the viscoelastic parameter, i.e. $\varepsilon De^2$. This suggests that the wetting time of the viscoelastic fluid increases with an increment in the viscoelastic parameter, as has been observed in Fig. 7(a).

We further make an effort in Fig. 7(b) to compare the results obtained from our analytical model with those obtained from the analytical model derived by Iwamatsu.[5] This comparison is particularly shown for the case of *a Newtonian fluid*. In the present study, the value of the viscoelastic parameter $De = 0$, and in the referred study of Iwamatsu[5], the value of the power-law index equal to one $(n = 1)$, both signify the case of a Newtonian fluid. From Fig. 7(b), it is observed that the present analytical model successfully verifies the theoretical results of Iswamatsu [5]. Note that the spreading behaviour of a Newtonian drop on a spherical substrate is discussed in the reported paper.[5]

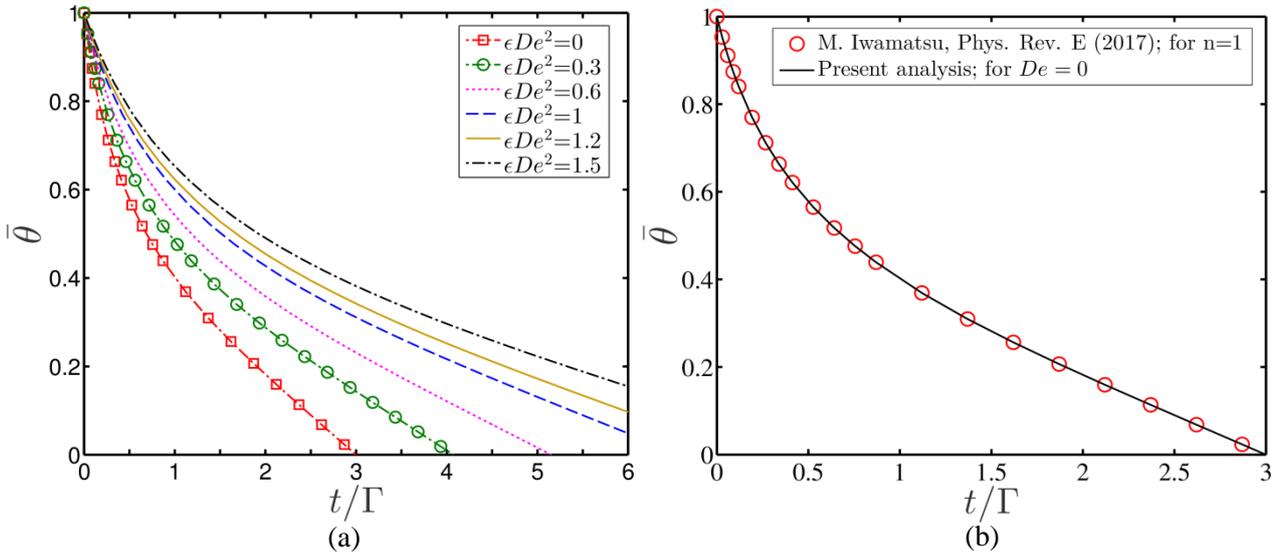

**Figure 7** (**will appear colour online**): (a) Plot showing the temporal variation of normalised contact angle $\bar{\theta}$ (normalised with the initial contact angle $\theta_0 = 1$) for different values of $\varepsilon De^2$. With increasing viscoelasticity of the fluid as realised by the increasing value of $\varepsilon De^2$, the complete wetting time of the surface becomes higher. (b) Plot showing the temporal variation of contact angle (normalised with initial contact angle $\theta_0 = 1$). The solid line represents the variation obtained from the present analysis in the limiting case of $De = 0$, while the 'O' markers are used to indicate the results reported by Iswamatsu.[5]

Notable findings of the present endeavour include the development of a model for studying elastic effect modulated spreading dynamics of a non-Newtonian drop on the spherical substrate and subsequent benchmarking of the model with the experimental results. This article will fill a gap on '*the unavailability of experimental investigations of drop spreading on a spherical substrate*' still affecting the existing literature in this paradigm.

**C. Spreading dynamics: Flat substrate versus spherical substrate**

It is important to mention here that, for the spreading phenomenon on a flat surface, the radius of the three-phase contact line expands to infinity. Because of this, the line tension effect becomes unimportant for the underlying spreading on a flat substrate, and its magnitude becomes negative, as mentioned in the referred literature.[21] On the contrary, for the spreading phenomenon on the spherical substrate, the radius of contact line shrinks, and the curvature diverges (see Fig. 2(a)). Owing to this effect, to achieve the complete wetting stage, the magnitude of line tension must be positive for a spherical substrate.[6] It may be mentioned in this context here that, for the Newtonian liquid drop, the spreading exponent is 1/10 for flat surface (tanner law) and 1/2 for the spherical substrate.[1,4,5,47] Also, as reported in the literature, the dynamic contact angle and spreading exponent are co-related as $\theta \propto t^{-\alpha}$, where $\alpha$ is the spreading exponent. Since $\alpha_{sphere} > \alpha_{flat}$, the spreading rate for the Newtonian liquid drop is expected to be faster on the spherical substrate as compared to that on a flat substrate. However, due to the effect of fluid elasticity as considered in this analysis, we show that the spreading rate of a viscoelastic liquid drop, in sharp contrast to that of the Newtonian liquid drop, will decrease on both the surfaces. In particular, we demonstrate in this analysis that the spreading rate of viscoelastic drop on the spherical surface will decrease for the higher viscoelastic behaviour of the fluid. We attribute this phenomenon to the effect of the shear-thinning nature of the fluid, which becomes stronger with increasing the magnitude of the viscoelasticity parameter.

In the present study, the spreading law for the viscoelastic fluid can be obtained using Eq. (36). For the sake of completeness in the analysis, we rewrite this equation as follows:

$$\Gamma\left(\frac{1}{3} + \frac{2\varepsilon De^2}{5}\right)\frac{d\theta}{dt} + \theta\left[\frac{1}{2}\left(\theta^2 - \theta_Y^2\right) + \frac{r_0 - R}{r_0 \theta}\tilde{\lambda}\right] = 0 \qquad (43)$$

For completely wetting stage, the equilibrium contact angle $\theta_Y$ and the line tension $\tilde{\lambda}$ can be set to zero. Therefore, Eq. (43) reduces to,

$$\Gamma\left(\frac{1}{3} + \frac{2\varepsilon De^2}{5}\right)\frac{d\theta}{dt} + \frac{\theta^3}{2} = 0 \qquad (44)$$

Now using the scaling analysis, Eq. (44) can now be reduced to the following form as:

$$\theta \sim t^{-1/2}\left[\Gamma\left(\frac{1}{3} + \frac{2\varepsilon De^2}{5}\right)\right]^{1/2} \qquad (45)$$

It can be observed from the above proportionality law [Eq. (45)] that there exists an additional prefactor in the spreading law, solely depends on the rheology of the viscoelastic fluid. By setting this prefactor to zero, we can obtain the spreading law for the Newtonian fluid as given in Eq. (39).

**IV. Conclusion**

We have developed a hydrodynamic model, consistent with the energy balance approach, to predict the spreading dynamics of a viscoelastic liquid drop on a spherical substrate. To represent the rheology of the viscoelastic fluid, we have considered a simplified Phan–Thien–Tanner model. By comparing the theoretical predictions with our experimental results, we have established that the present modelling framework can successfully predict the phenomenon in the late-stage of the spreading. To calculate the dissipation at the contact line, we have modelled the drop spreading to be a shrinking crater on the flat substrate and have considered the effect of the line tension. Considering the applicability of our model in describing the phenomenon during the late-stage of spreading on a spherical substrate, the magnitude of line tension, in sharp contrast to its negative value for the spreading on a flat surface, is taken as positive necessarily to achieve the complete spreading. We have unveiled that the velocity in the drop flow field, which can be well-approximated to the contact line velocity in the capillary driven regime, shows similar qualitative behaviour with our experimental observations. Also, we have established the expression of the dynamic contact angle for the spreading of a viscoelastic drop on a spherical substrate compatible with the capillary-viscous forcing regime. We have shown that the elastic nature of viscoelastic fluid leads to an increment in the dynamic contact angle at any temporal instant as compared to its Newtonian counterpart. Finally, we have shown that the phenomenon of increasing contact angle results in the enhancement in the complete wetting time of viscoelastic drop having higher viscoelasticity. We believe that the present analysis may have far-ranging consequences towards the long-awaited demand of experimental investigations of the spreading of the elastic non-Newtonian drop on a spherical substrate.

**Acknowledgment**

PKM acknowledges the financial grants obtained from DST-SERB through project No ECR/2016/000702/ES. Authors acknowledge Microfluidics Laboratory, IIT Guwahati for the experimental facilities.

**For Table of Contents Use Only**

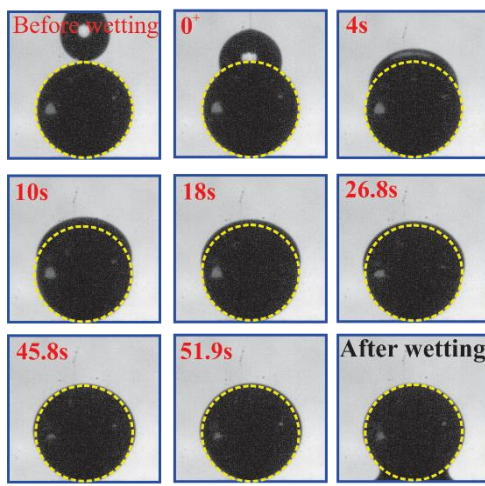
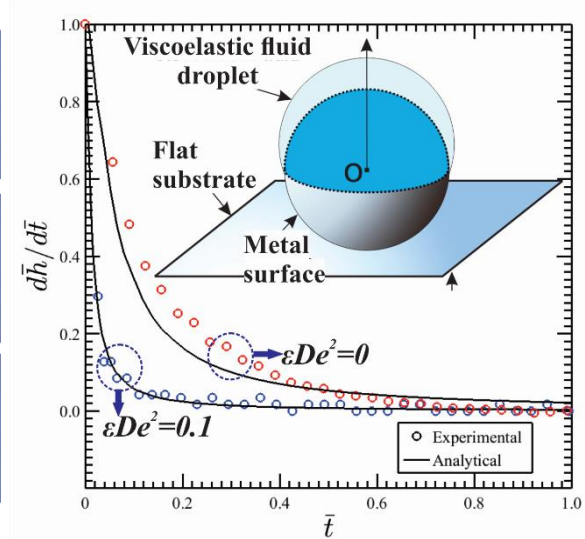